\documentclass[aps,twocolumn,prb,showpacs,superscriptaddress]{revtex4-2}
\usepackage{makeidx}
\usepackage{amsfonts}
\usepackage{graphics}
\usepackage{graphicx}
\usepackage{amssymb}
\usepackage{amsthm}
\usepackage{amsmath}
\usepackage{hyperref}

\begin{document}

\title{Effect of spin-orbit coupling on spin and orbital ordering in Sr$_{n+1}$Cr$_n$O$_{3n+1}$, $n=1,2$}

\author{Cristian Fanjul}
\affiliation{Centro At\'{o}mico Bariloche and Instituto Balseiro, 8400 Bariloche, Argentina}

\author{A. A. Aligia}
\affiliation{Centro At\'{o}mico Bariloche and Instituto Balseiro, 8400 Bariloche, Argentina}

\affiliation{Instituto de Nanociencia y Nanotecnolog\'{\i}a
CNEA-CONICET, GAIDI,
Centro At\'{o}mico Bariloche, 8400 Bariloche, Argentina}

\date{\today }

\begin{abstract}
We incorporate spin-orbit coupling (SOC) into effective Kugel-Khomskii models for the $n=1$ and $n=2$ members of the Ruddlesden-Popper series
Sr$_{n+1}$Cr$_n$O$_{3n+1}$. These model contain
interacting spins 1 and pseudospins 1/2 at each site describing spin and 
orbitals degrees of freedom respectively. 
We solve the models at zero temperature using pseudospin bond operators
and spin waves.
We find that for realistic parameters, SOC dominates the physics of
the compound Sr$_{2}$CrO$_{4}$ with almost decoupled single CrO$_2$ planes.
The spin ordering is antiferromagnetic, with nearest-neighbor Cr spins aligned antiparallel. The corresponding orbital configuration is
$d_{xy \uparrow }^{1}(d_{xz\uparrow }^{1}-id_{yz\uparrow }^{1})$
or
$d_{xy \downarrow }^{1}(d_{xz\downarrow }^{1}+id_{yz\downarrow }^{1})$
depending on the spin of the site.
In contrast, for the bilayer compound Sr$_{3}$Cr$_{2}$O$_{7}$ we find
that the effect of the SOC is weak and the system prefers to form pseudospin
singlets in the $z$ direction perpendicular to the planes.
The spin order is antiferromagnetic within each plane and ferromagnetic between planes, in agreement with previous studies.

\end{abstract}

\maketitle

\section{Introduction}
\label{intro}

For decades, the long-range ordering of orbitals and its intricate coupling with spin order have been a subject of significant scientific interest. This interplay was established in a seminal theoretical study by Kugel and Khomskii, who examined the entangled orbital and spin degrees of freedom in the model compounds KCuF$_3$ and K$_2$CuF$_4$ \cite{Kugel82}.
The staggering ordering of the $e_g$ orbitals ($x^2-y^2$ or $3z^2-r^2$) results in a corresponding staggered pattern of quadrupolar distortions within the CuF$_4$ units across the
$ab$-planes, as expected from  electron-phonon
interaction \cite{Liech95}.
Other examples of compounds demonstrating orbital order can be found in Refs.
\cite{Mizo95,Feiner97,Ul03,Ali04,Sugai06,Lee06,Ray07,Manaka07,Normand08,
Kata09,Stin11,Nun13,Lobos14,Gio14,Chiz25,
Ortega07,Lee09,Carta22,Doyle24,Meier26,Ishi17,Jean19,Pandey21,Moha23,Jean17,Ali19}.

In particular in the Ruddlesden-Popper chromates Sr$_{n+1}$Cr$_n$O$_{3n+1}$, the evolution with $n$ has been studied recently, displaying increased metallicity
with increasing $n$ \cite{Doyle24}. Also the possibility of (anti-)altermagnetism
has been addressed \cite{Meier26}. The end of the series $n \rightarrow \infty$ corresponds
to SrCrO$_3$, which had been studied previously \cite{Ortega07,Lee09,Carta22}.
Other specific compounds studied include the $n=1$ member Sr$_2$CrO$_4$ \cite{Sugi14,Ishi17,Jean19,Pandey21,Moha23} as well as the $n=2$ member
Sr$_{3}$Cr$_{2}$O$_{7}$ \cite{Jean17,Ali19}. These systems are characterized by the presence of Cr$^{4+}$, each possessing two electrons in the $t_{2g}$ manifold.
While this local electronic configuration is common, distinct spin and orbital ordered phases emerge in different layered structures. In the single layer compound Sr$_2$CrO$_4$ there is evidence of antiferromagnetic (AF) spin ordering  \cite{Sugi14,Ishi17,Jean19,Pandey21,Moha23} and theoretical
results that favor AF orbital ordering  \cite{Ishi17,Yama19,Pandey21,Moha23} in absence of spin-orbit coupling (SOC).
For the related insulating compound BaCrO$_3$, calculations using density-functional theory (DFT) and dynamical mean-field theory lead to AF spin and orbital ordering in the CrO$_2$ planes \cite{Gio14}.

In the bilayer compound Sr$_3$Cr$_2$O$_7$,
the magnetic structure observed by neutron diffraction is consistent with
results of DFT \cite{Jean19} and corresponds to AF alignment between nearest-neighbors within the $x,y$ planes and ferromagnetic (FM)
between planes ($z$ direction).
There is a magnetic transition at 210 K with a huge total entropy change near
$R\ln(6)$ indicating a simultaneous spin and orbital ordering.
Orbital ordering is usually not detected in DFT
due to the difficulties of these techniques to obtain orbital
polarization \cite{Liech95,Ali13}. However, calculations suggest the presence of interplane orbital singlets \cite{Jean17,Ali19}.
A similar physics was observed in K$_3$Cr$_2$O$_7$, where distortions reveal AF orbital ordering, while the system presents a spin gap due to  spin dimers in the $z$ direction \cite{Manaka07}.

In the above discussion, as well as in most of the research papers, SOC has been neglected. However, recently, for the one-layer compound,
Mohapatra {\it et al.}
\cite{Moha23}
found a transition to entangled orbital order at a moderate
value of the SOC $\lambda$. The authors used a Hubbard-Kanamori model solved in
a self-consistent mean-field approximation and including fluctuations.
An alternative approach used previously \cite{Ali19} is to use a Kugel-Khomskii Hamiltonian (KKH) that describes the spin and orbital degrees of freedom constructed
by perturbation theory from the three-band Hubbard-Kanamori model, whose
hoppings were determined from ab-initio calculations. By treating the strong Coulomb interactions exactly, this approach avoids common mean-field approximations and provides a more reliable starting point for the quantum many-body description.

The goal of this work is to investigate the effect of SOC in both compounds
Sr$_2$CrO$_4$ and Sr$_3$Cr$_2$O$_7$. We solve
the corresponding KKHs using spin- and pseudospin-wave theory
For the second compound we also use the
bond-operator formalism (BOF) \cite{Gopa94,Matsu04,Nor11,Muniz14}, which is more appropriate to treat the orbital dimers in the
$z$ direction.

The manuscript is structured as follows. Section~\ref{model} introduces the effective Kugel-Khomskii Hamiltonians (KKHs) used to model the two compounds. The resulting physical properties for realistic parameters are presented in Section~\ref{res}. Finally, Section~\ref{sum} provides a comprehensive summary and discussion of our findings. A detailed derivation of a central, recurrent calculation is contained in the Appendix.

\section{The models}

\label{model}

We study two KKHs to describe the spin and orbital degrees of freedom, one
that contains one CrO$_{2}$ plane in the $x,y$ directions to describe
Sr$_{2} $CrO$_{4}$ and the other contains two CrO$_{2}$ planes coupled in the
$z $\ direction to describe Sr$_{3}$Cr$_{2}$O$_{7}$. The Hamiltonian for the
latter but without SOC has been derived in Ref. \cite{Ali19}. The authors
constructed first a three-band Hubbard model for effective Cr $t_{2g}$
orbitals (containing some admixture with O orbitals) that were derived from
maximally localized Wannier functions. Then, the KKH was obtained by
perturbation theory. We take this Hamiltonian and its parameters derived in Ref. \cite{Ali19} as our starting
point. Smaller couplings in the $z$ direction across SrO layers are neglected. Therefore the
Hamiltonians are restricted to two dimensions.

The Hamiltonian for the bilayer system can be written as

\begin{equation}
H=H_{1}+H_{2}+H_{c},  \label{ham}
\end{equation}
where $H_{i}$ describes the CrO$_{2}$ plane $i$ and $H_{c}$ corresponds to
the coupling between both planes. Clearly, the monolayer system is described
by $H_{i}$, which in turn is split as

\begin{equation}
H_{i}=H_{i}^{0}+H_{i}^{\text{SOC}},  \label{hi}
\end{equation}
where $H_{i}^{\text{SOC}}$ is the SOC extended to the plane $i$ and
$H_{i}^{0}$, $H_{c}$ contain the terms described before \cite{Ali19}
that we explain below.

There are two low-energy configurations for the Cr$^{4+}$ ions. We use a
pseudospin $\mathbf{T}_{i\mathbf{r}}$, with $T_{i\mathbf{r}}^{z}=-1/2$ (1/2)
to denote the $d_{xy}^{1}d_{xz}^{1}$ ($d_{xy}^{1}d_{yz}^{1}$) configuration
of plane $i$ at the two-dimensional position $\mathbf{r}$ within the plane.
The spin 1 vector at each site (imposed by Hund rules) is denoted by
$\mathbf{S}_{i\mathbf{r}}$. With this notation one has \cite{Ali19}

\begin{eqnarray}
H_{i}^{0} &=&\sum_{\mathbf{ra}}[\frac{I_{S}^{p}}{4}
\mathbf{S}_{i\mathbf{r}}\cdot \mathbf{S}_{i\mathbf{r}
+\mathbf{a}}+I_{T}^{p}T_{i\mathbf{r}}^{z}T_{i \mathbf{r}+\mathbf{a}}^{z}  \notag \\
&&+I_{ST}^{p}(\mathbf{S}_{i\mathbf{r}}\cdot \mathbf{S}_{i\mathbf{r}+
\mathbf{a}})T_{i\mathbf{r}}^{z}T_{i\mathbf{r}+\mathbf{a}}^{z}]  \notag \\
&&+I_{A}\sum_{\mathbf{r}}[-(\mathbf{S}_{i\mathbf{r}}\cdot
\mathbf{S}_{i\mathbf{r+}a\mathbf{\hat{x}}})(T_{i\mathbf{r}}^{z}
+T_{i\mathbf{r+}a\mathbf{\hat{x}}}^{z})  \notag \\
&&+(\mathbf{S}_{i\mathbf{r}}\cdot
\mathbf{S}_{i\mathbf{r+}a\mathbf{\hat{y}}})
(T_{i\mathbf{r}}^{z}+T_{i\mathbf{r+}a\mathbf{\hat{y}}}^{z})],  \label{hp}
\end{eqnarray}
where $a$ is the lattice parameter and $\mathbf{a=}a\mathbf{\hat{x}}$ or
$a \mathbf{\hat{y}}$ are two independent vectors connecting nearest-neighbor Cr
ions in the plane. The factor 1/4 in the first term was chosen to compensate
the difference between the magnitude of the spin $S=1$ and the pseudospin $T=1/2$ to facilitate comparison
between different terms.
For the remainder of this work, we anticipate that the anisotropic term proportional to $I_{A}$ does not play any role. The spin interactions are AF along both directions, so the effective field acting on each pseudospin $T_{i\mathbf{r}}^{z}$ cancels out.

The coupling between planes has the form

\begin{eqnarray}
H_{c} &=&\sum_{\mathbf{r}}[\frac{I_{S}}{4}\mathbf{S}_{1\mathbf{r}}\cdot
\mathbf{S}_{2\mathbf{r}}+I_{T}\mathbf{T}_{1\mathbf{r}}\cdot
\mathbf{T}_{2 \mathbf{r}}  \notag \\
&&+I_{ST}(\mathbf{S}_{1\mathbf{r}}\cdot \mathbf{S}_{2\mathbf{r}})
(\mathbf{T}_{1\mathbf{r}}\cdot \mathbf{T}_{2\mathbf{r}})].  \label{hc}
\end{eqnarray}

For any transition metal ion, the SOC has the form
$H_{\text{ion}}^{\text{SOC}}=\lambda \sum_{j}\mathbf{l}_{j}\cdot \mathbf{s}_{j}$, where $\mathbf{l}_{j}$ ($\mathbf{s}_{j}$) is the orbital angular momentum
(spin) of
each electron. In presence of crystal field, in many cases the effect of
$H_{\text{ion}}^{\text{SOC}}$ is of second order and can be neglected.
However, in our case, the degeneracy of $xz$ and $yz$ orbitals renders
 the contribution $\sum_{i}\,l_{i}^{z}\,s_{i}^{z}$\ relevant at
first order ($l_{i}^{+}$ and $l_{i}^{-}$ mix $t_{2g}$ with $e_{g}$ states
and can be neglected at first order). In terms of creation and annihilation
of operators for the d $t_{2g}$ orbitals one has for the relevant part of
$H_{\text{ion}}^{\text{SOC}}$

\begin{eqnarray}
\sum_{j}\,l_{j}^{z}\,s_{j}^{z} &\simeq &\frac{i}{2}(d_{yz\uparrow }^{\dagger
}d_{xz\uparrow }-d_{xz\uparrow }^{\dagger }d_{yz\uparrow }  \notag \\
&&-d_{yz\downarrow }^{\dagger }d_{xz\downarrow }+d_{xz\downarrow }^{\dagger
}d_{yz\downarrow }),  \label{soc1}
\end{eqnarray}
where $i$ is the imaginary unit. Taking into account that for our system,
the spin of the $d_{xy}$ orbital is the same as the other one, using the
same spin and pseudospin operators as above, and extending the result to all
atoms in the plane one has

\begin{equation}
H_{i}^{\text{SOC}}=-\lambda \sum_{\mathbf{r}}S_{i\mathbf{r}}^{z}
T_{i\mathbf{r}}^{y}.  \label{soc2}
\end{equation}

This completes the description of the Hamiltonian.

To estimate the value of $\lambda $ we recall that within each atomic term
(in particular $^{3}F$ of Cr$^{4+}$) $H_{\text{ion}}^{\text{SOC}}=\Lambda
\mathbf{L}\cdot \mathbf{S}$, where $\mathbf{L=}\sum_{j}\mathbf{l}_{j}$,
$\mathbf{S=}\sum_{j}\mathbf{s}_{j}$, and $\Lambda =\lambda /(2S)$. From the
known atomic energy levels of Cr$^{4+}$ \cite{Moore} and the difference
between energies $E(J)$ with different $J$ ($\mathbf{J=L}+\mathbf{S}$),
$E(J)-E(J-1)=\Lambda J$, one can estimate $\Lambda =20$ meV within an error
of 6\%. Therefore $\lambda \simeq 40$ meV. The parameters of $H_{i}^{0}$ and
$H_{c}$ have been estimated before \cite{Ali19}. The values of the
parameters used corresponding to atomic Hund rules exchange $J_{H}=0.7$ eV
\cite{Ali13} and Coulomb repulsion $U=4.1$ eV are given in Table \ref{tab1}

\begin{table}[h]
\caption{Parameters of $H$ in meV}
\label{tab1}
\begin{ruledtabular}
\begin{tabular}{llllllll}
$I_S$ & $I_T$ & $I_{ST}$ & $I_{S}^{p}$ & $I_{T}^{p}$ & $I_{ST}^{p}$ & $I_{A}$ & $\lambda$
\\
\hline
4.2 & 68.1 & 42.4 & 54.7 & 28.2 & 17.6 & 9.7 & 40 \\

\end{tabular}
\end{ruledtabular}
\end{table}

The dominant interaction is the interplane vertical one (in the $z$-direction)
for the bilayer system  $I_T$, which favours pseudospin
singlets, or possibly AF pseudospin vertical order
(corresponding to different orbital configurations in both planes).
This is
due to the fact that the strongest hopping between $xz,yz$ Cr orbitals is in
the $z$ direction, perpendicular to the planes \cite{Ali19} and both
orbitals can hop, while in the $x$ direction for example, by symmetry, only
the $xz$ orbital can hop to the nearest-neighbor Cr atom, through an
intermediate O $p_{z}$ orbital. Similarly, while the $xy$ orbital can hop in
both directions in the plane, it cannot hop in the $z$ direction through
intermediate O orbitals.
This explains the weak AF spin
interaction in the $z$ direction.
This results in a weak AF spin interaction along $z$.
Combined with the effect of $I_{ST}$, which is approximately an order of magnitude larger than $I_S$, this weak AF coupling is overcome,
and ferromagnetic spin alignment between planes is clearly favored.
Instead, in the planes the dominant interaction is $I_{S}^{p}$ which favors spin AF order. In addition the interaction between
spins and pseudospins $I_{ST}^{p}$ is smaller than
$I_{T}^{p}$ and therefore
AF orbital order in the plane is also expected.

A detailed study of the competition between vertical pseudospin singlets and
long-range pseudospin AF ordering in presence of SOC is the subject of Section
\ref{two}.

\section{Results}

\label{res}

For the compound Sr$_{2}$CrO$_{4}$,  in which the CrO$_{2}$ planes are
separated by two planes of SrO and are therefore relatively isolated, we solve the
single-plane Hamiltonian $H_{i}$ using spin- and pseudospin-wave theory. For
Sr$_{3}$Cr$_{2}$O$_{7}$, in which two CrO$_{2}$ planes in the $x,y$ directions are
strongly coupled along the $z$ direction,
quantum fluctuations in this direction within the pseudospin sector can be crucial, leading to the formation of singlet orbital dimers. To describe the phase of these singlets (which we denote as phase I), pseudospin-wave theory is not appropriate; instead, we employ the
BOF \cite{Gopa94,Matsu04,Nor11,Muniz14} in the form
given in Ref. \cite{Muniz14}, starting from the singlet ground state of the pseudospin dimers in the $z$ direction.
The alternative scenario (phase II) corresponds to a state in which the expectation value of the pseudospin at each site is nonzero and can be described using pseudospin-wave theory.
In Section \ref{two} we compare the energy of both phases.

In general, except for pseudospin singlets in the BOF, the classical energy of the spins and pseudospins is minimised
determining the orientation of these quantities, and then fluctuations are
introduced, retaining only quadratic terms in the bosonic operators. The
linear terms vanish as a consequence of the minimization of the classical
energy (see appendix A of Ref. \cite{Henry12}) and terms of order higher
than two are neglected. The problem is separated into different ones for the
spin and pseudospin sector, in which classical values (or those
corresponding to the pseudospin singlet if the BOF is used) of one sector
act as effective field for the other one. Specifically for a product of an
operator $O(S)$ of a spin and $O(T)$ of a pseudospin we approximate

\begin{eqnarray}
O(S)O(T) &\simeq &\left\langle O(S)\right\rangle _{c}O(T)+O(S)\left\langle
O(T)\right\rangle _{c}  \notag \\
&&-\left\langle O(S)\right\rangle _{c}\left\langle O(T)\right\rangle _{c},
\label{dec}
\end{eqnarray}
where $\left\langle O(S)\right\rangle _{c}$ is the classical value of the
spin operator and $\left\langle O(T)\right\rangle _{c}$ is either the
classical value of the pseudospin operator or the expectation value if the
BOF is used. The first (second) term of the second member takes part of the
pseudospin (spin) sector of the problem and the third term avoids double
counting of it in the total energy.

\subsection{ One plane}
\label{one}

From the magnitude of the interactions in Table \ref{tab1}, it is clear that
in absence of SOC ($\lambda =0$) AF
order of both spins and
pseudospins is favoured, with nearest-neighbour spins and pseudospin pointing
in opposite $\pm z$ directions, in agreement with previous results
\cite{Pandey21,Moha23}. For the following discussion, it is convenient to perform
a rotation in half of the sites an angle $\pi $ around the $x$ axis so that
all spins and pseudospins point in the same direction for $\lambda =0$ and
the problem has full translational symmetry. The new operators are denoted
with a tilde. After this transformation, the problem in the pseudospin
sector from Eqs. (\ref{hi}), (\ref{hp}), (\ref{soc2}) and (\ref{dec}) with
$\left\langle \mathbf{\tilde{S}}_{i\mathbf{r}}\cdot
\mathbf{\tilde{S}}_{i\mathbf{r}+\mathbf{a}}\right\rangle _{c}=-\left\langle \mathbf{S}_{i\mathbf{r}}\cdot
\mathbf{S}_{i\mathbf{r}+\mathbf{a}}\right\rangle _{c}=1$,
$\left\langle \tilde{S}_{i\mathbf{r}}^{z}\right\rangle _{c}=1$, dropping the
subscript $i$ for simplicity, takes the form of a transverse Ising model in
two dimensions \cite{Friedman78,Blote02,Schmitt22}

\begin{eqnarray}
H_{1T} &=&-\tilde{I}_{T}^{p}\sum_{\mathbf{ra}}
\tilde{T}_{\mathbf{r}}^{z}\tilde{T}_{\mathbf{r}+\mathbf{a}}^{z}-B_{T}\sum_{\mathbf{ra}}\tilde{T}_{\mathbf{r}}^{y},  \notag \\
\tilde{I}_{T}^{p} &=&I_{T}^{p}-I_{ST}^{p},\text{ \ }B_{T}=\lambda .
\label{htp}
\end{eqnarray}
The model has a quantum phase transition at a critical value $B_{c}$, where
the gap vanishes. For $B_{T}<B_{c}$, the expectation value
$\langle \tilde{T}_{\mathbf{r}}^{z}\rangle \neq 0$, while
$\langle \tilde{T}_{\mathbf{r}}^{z}\rangle =0$ for $B_{T}\geqslant B_{c}$. From cluster Monte Carlo
simulations, $B_{c}=1.52219 \tilde{I}_{T}^{p}$ \cite{Blote02},
while in our classical approach $B_{c}=2\tilde{I}_{T}^{p}$.
Using the values of Table \ref{tab1},
$\tilde{I}_{T}^{p}=10.6$ meV and $B_{T}=40$ meV.
Therefore the system is well inside the phase dominated by the SOC, in which
the orbital occupancy is dominated by either
$d_{xy\uparrow }^{\dagger }(d_{xz\uparrow }^{\dagger}
-id_{yz\uparrow }^{\dagger }) |0 \rangle$ or
$d_{xy\downarrow }^{\dagger}(d_{xz\downarrow }^{\dagger}
+id_{yz\downarrow }^{\dagger })|0 \rangle$
depending on the value of
$S_{\mathbf{r}}^{z}$ at the corresponding site.
Following Ref. \cite{Moha23} we denote this phase as entangled orbital phase.

Similarly, in the spin sector, using $\left\langle
T_{i\mathbf{r}}^{z}T_{i\mathbf{r}+\mathbf{a}}^{z}\right\rangle _{c}=0$,
$\langle T_{\mathbf{r}}^{y}\rangle _{c}=1/2$ corresponding to the
classical values for $B_{T}>B_{c} $, the Hamiltonian takes the form

\begin{eqnarray}
H_{1S} &=&\frac{I_{S}^{p}}{4}\sum_{\mathbf{ra}}
\left( \frac{\tilde{S}_{\mathbf{r}}^{+}\tilde{S}_{\mathbf{r}
+\mathbf{a}}^{+}+\tilde{S}_{\mathbf{r}}^{-}\tilde{S}_{\mathbf{r}+\mathbf{a}}^{-}}{2}
-\tilde{S}_{\mathbf{r}}^{z}\tilde{S}_{\mathbf{r}+\mathbf{a}}^{z}\right)  \notag \\
&&-B_{S}\sum_{\mathbf{ra}}\tilde{S}_{\mathbf{r}}^{z},\text{ \ }B_{S}=\frac{\lambda }{2}.  \label{hsp}
\end{eqnarray}
To solve this Hamiltonian, it is usual to introduce bosonic operators by
means of a Holstein-Primakoff transformation

\begin{equation}
\tilde{S}_{\mathbf{r}}^{+}\simeq \sqrt{2S}b_{\mathbf{r}},
\text{ }\tilde{S}_{\mathbf{r}}^{-}\simeq \sqrt{2S}b_{\mathbf{r}}^{\dagger },\text{ }\tilde{S}_{\mathbf{r}}^{z}=S-b_{\mathbf{r}}^{\dagger }b_{\mathbf{r}}.  \label{hol}
\end{equation}
for $S=1$. Fourier transforming $b_{\mathbf{r}}=\sqrt{1/N}\Sigma
_{k}e^{-ik\cdot \mathbf{r}}b_{k}$, where $k$ is a two-dimensional wave
vector of the square lattice, and neglecting terms of fourth order in the
bosonic operators, $H_{S}$ takes the form of an exactly solvable quadratic
bosonic problem

\begin{equation}
H_{b}=C+\sum_{k}\left[ \omega _{0k}b_{k}^{\dagger }b_{k}+\left( \frac{\omega
_{1k}}{2}b_{k}^{\dagger }b_{-k}^{\dagger }+\text{H.c.}\right) \right] ,
\label{hb}
\end{equation}
with $C=-N(B_{S}+I_{S}^{p}/2)$ where $N$ is the number of Cr atoms in
the plane, $\omega _{0k}=4 I_{S}^{p}+B_{S}$, and
$\omega _{1k}=I_{S}^{p}\gamma _{k}$, with

\begin{equation}
\gamma _{k}=\frac{1}{2}\left( \cos k_{x}+\cos k_{y}\right) .  \label{gk}
\end{equation}
For completeness and since $H_{b}$ appears several times, its solution is
included in the appendix. From there, the expectation value of the local
spin projection including quantum corrections is

\begin{equation}
\left\langle \text{ }\tilde{S}_{\mathbf{r}}^{z}\right\rangle =
1-\frac{1}{N}\sum\limits_{k}\left\langle b_{k}^{\dagger }b_{k}\right\rangle =1-\frac{1}{N}\sum\limits_{k}|v_{k}|^{2},  \label{sz}
\end{equation}
and the ground state energy of $H_{b}$ is given by Eq. (\ref{ene}).

For the pseudospin problem we introduce bosons in a similar way as Eq.
(\ref{hol}) but with $S$ replaced by the magnitude of the pseudospin $T=1/2$, and
the axis rotated 120 degrees so that $z\longrightarrow y\longrightarrow
x\longrightarrow z$, and then $\tilde{T}_{\mathbf{r}}^{x}
=(\tilde{T}_{\mathbf{r}}^{+}+\tilde{T}_{\mathbf{r}}^{-})/2\longrightarrow
\tilde{T}_{\mathbf{r}}^{z}$. Specifically

\begin{equation}
\tilde{T}_{\mathbf{r}}^{y}=\frac{1}{2}-a_{\mathbf{r}}^{\dagger }a_{\mathbf{r}%
},\text{ }\tilde{T}_{\mathbf{r}}^{z}=\frac{a_{\mathbf{r}}^{\dagger }+a_{%
\mathbf{r}}}{2}. \label{holt}
\end{equation}
Fourier transforming $a_{\mathbf{r}}=\sqrt{1/N}\Sigma _{k}e^{-ik\cdot
\mathbf{r}}a_{k}$, the Hamiltonian in the pseudospin sector becomes

\begin{eqnarray}
H_{1T} &=&-\frac{\tilde{I}_{T}^{p}}{2}\sum\limits_{k}\gamma _{k}\left(
2a_{k}^{\dagger }a_{k}+a_{k}^{\dagger }a_{-k}^{\dagger }+a_{k}a_{-k}\right)
\notag \\
&&-\frac{1}{2}NB_{T}+B_{T}\sum\limits_{k}a_{k}^{\dagger }a_{k},
\label{htb}
\end{eqnarray}
which also has the form of Eq. (\ref{hb}), with $C=-NB_{T}/2$, $\omega
_{0k}=B_{T}-\tilde{I}_{T}^{p}\gamma _{k}$ and
$\omega _{1k}=\tilde{I}_{T}^{p}\gamma _{k}$. Using the results of the appendix one obtains for the
correlation functions including quantum fluctuations

\begin{eqnarray}
\left\langle \tilde{T}_{\mathbf{r}}^{y}\right\rangle &=&\frac{1}{2}
-\frac{1}{N}\sum\limits_{k}v_{k}^{2},\text{ }  \notag \\
\left\langle \tilde{T}_{\mathbf{r}}^{z}\tilde{T}_{\mathbf{r}
+\mathbf{a}}^{z}\right\rangle &=&\frac{1}{4N}\sum\limits_{k}\gamma
_{k}(u_{k}+v_{k})^{2}.  \label{corr}
\end{eqnarray}

For the parameters of Table \ref{tab1} we obtain
$\left\langle \text{ }\tilde{S}_{\mathbf{r}}^{z}\right\rangle =0.9972$
(almost saturated compared to the value 0.805 in absence of SOC), $\left\langle \tilde{T}_{\mathbf{r}}^{y}\right\rangle =0.4989,\left\langle T_{\mathbf{r}}^{z}T_{\mathbf{r}+ \mathbf{a}}^{z}\right\rangle
=-\left\langle \tilde{T}_{\mathbf{r}}^{z}\tilde{T}_{\mathbf{r}+\mathbf{a}}^{z}\right\rangle = -0.0085$.

The AF spin order is in agreement with experiment \cite{Sugi14}.

If AF orbital ordering were present (as expected for small $\lambda$ as discussed above), one would anticipate a corresponding distortion of the atomic structure, reflecting the asymmetry in the hopping amplitudes.
To illustrate this, consider an O$^{2-}$ ion located between two nearest-neighbor
Cr$^{4+}$ ions along the $x$ direction, with the left Cr site occupied by a $xz$ electron and the right one by a $yz$ electron.
By symmetry, the only allowed hopping between the O $p$ orbitals and the above mentioned Cr orbitals
is between the O $p_z$ orbital and the Cr $xz$ orbital.
As a result, both $p_z$ electrons can hop to the right, whereas
by Pauli principle, only the electron with spin opposite to that of the $xz$ electron can hop to the left.
A displacement of the O atom to the right enhances (reduces) the O–Cr hopping amplitude $t_p$ to the Cr atom at the right (left) Cr ion. Owing to the resulting asymmetry, such a displacement is energetically favorable.
However, no such lattice distortions have been observed experimentally, which is consistent with our results.

\subsection{ Two planes}
\label{two}

In Sr$_{3}$Cr$_{2}$O$_{7}$, as it is evident from Table \ref{tab1}, the
strongest interaction is the pseudospin one between planes $I_{T}$, while $I_S$ is very weak. Thus, in order to take advantage of the
last term in Eq. (\ref{hc})
$I_{ST}(\mathbf{S}_{1\mathbf{r}}\cdot \mathbf{S}_{2\mathbf{r}})
(\mathbf{T}_{1\mathbf{r}}\cdot \mathbf{T}_{2\mathbf{r}})$, it
is convenient to order ferromagnetically the spins in the $z$ direction, as
experimentally observed and in agreement with ab initio calculations
\cite{Ali19}, while the spin order is AF in the planes. Therefore,
the effective pseudospin interaction in the $z$ direction is $I_{T}+I_{ST}$,
favouring strongly pseudospin singlet dimers in the $z$ direction, as found
in previous calculations without SOC \cite{Jean17,Ali19}. We denote by I
this phase with orbital singlets.
Owing to the energy gap between the pseudospin triplet and singlet states, the effect of SOC in this phase enters only at second order and is therefore small, as we show below.

In contrast, in the other possible phase denoted as II, $\left\langle
\mathbf{T}_{i\mathbf{r}}\right\rangle \neq 0$ and SOC has a first-order
contribution and might turn this phase into the favored one, if the spins
and pseudospins are aligned conveniently. In this Section we calculate the
energy of both phases and compare them.

\subsubsection{Phase I}

To solve the pseudospin sector of this phase, we use the bond-operator
formalism in the form of a generalized spin-wave theory \cite{Muniz14}. The
vertical pseudospin singlet $(|\uparrow \downarrow \rangle -|\downarrow
\uparrow \rangle )/\sqrt{2}$, where the first arrow denotes
$T_{1\mathbf{r}}^{z}$, is represented using a singlet boson operator as
$s_{\mathbf{r}}^{\dagger }|0\rangle $, where $|0\rangle $ represents the boson vacuum.
The three triplets are represented by $t_{\mathbf{r}\gamma }^{\dagger
}|0\rangle $. The number of triplet excitations is assumed small and the
singlets are "condensed" using

\begin{equation}
s_{\mathbf{r}}^{\dagger }=s_{\mathbf{r}}=\sqrt{1-\Sigma
_{\gamma }t_{\mathbf{r}\gamma }^{\dagger }t_{\mathbf{r}\gamma }}.
\label{cond}
\end{equation}
The intraplane term in the Hamiltonian mixes the singlet
$s_{\mathbf{r}}^{\dagger }|0\rangle $ with the triplet with projection 0, which we denote
by $\alpha $: $t_{\mathbf{r}\alpha }^{\dagger }|0\rangle =(|\uparrow
\downarrow \rangle +|\downarrow \uparrow \rangle )/\sqrt{2}$, because
$T_{1\mathbf{r}}^{z}s_{\mathbf{r}}^{\dagger }|0\rangle =t_{\mathbf{r}\alpha
}^{\dagger }|0\rangle /2$, $T_{2\mathbf{r}}^{z}s_{\mathbf{r}}^{\dagger
}|0\rangle =-t_{\mathbf{r}\alpha }^{\dagger }|0\rangle /2$, and the same
interchanging $s_{\mathbf{r}}$ and $t_{\mathbf{r}\alpha }^{\dagger }$. In
addition from the SOC term Eq. (\ref{soc2})

\begin{equation}
\sum\limits_{j=1}^{2}T_{j\mathbf{r}}^{y}t_{\mathbf{r}\alpha }^{\dagger
}|0\rangle =\frac{-i}{\sqrt{2}}(|\uparrow \uparrow \rangle -|\downarrow
\downarrow \rangle )=t_{\mathbf{r}\beta }^{\dagger }|0\rangle .
\label{tbeta}
\end{equation}
Similarly as before, we use the symmetry operation $R$, which rotates spins and
pseudospins in half of the sites $j\mathbf{r}$ an angle $\pi $ around the $x$
axis and in the same way for $j=1$ and 2, so that
the system has translational invariance in the plane.
We define
$\mathbf{\tilde{S}}_{j\mathbf{r}}=R\mathbf{S}_{j\mathbf{r}}R^{\dagger }$ and
$\mathbf{\tilde{T}}_{j\mathbf{r}}=R\mathbf{T}_{j\mathbf{r}}R^{\dagger }$.
Note that
$R$ does not affect the singlets, so that the triplet operators can be used
in the rotated basis.
Using these results and the decoupling
Eq. (\ref{dec}) with $\left\langle \mathbf{\tilde{S}}_{1\mathbf{r}}\cdot
\mathbf{\tilde{S}}_{2\mathbf{r}}\right\rangle _{c}=\left\langle
\mathbf{\tilde{S}}_{i\mathbf{r}}\cdot \mathbf{\tilde{S}}_{i\mathbf{r}
+\mathbf{a}}\right\rangle _{c}=\left\langle \tilde{S}_{i\mathbf{r}}^{z}\right\rangle
_{c}=1$, the Hamiltonian in the pseudospin sector takes the form

\begin{eqnarray}
H_{T}^{\text{I}} &=&-\frac{3N}{4}\tilde{I}_{T}+\tilde{I}_{T}
\sum_{\mathbf{r}\gamma }t_{\mathbf{r}\gamma }^{\dagger }t_{\mathbf{r}\gamma }  \notag \\
&&-\sum_{\mathbf{r,a}}\frac{\tilde{I}_{T}^{p}}{2}\left( t_{\mathbf{r}\alpha
}^{\dagger }t_{\mathbf{r}+\mathbf{a}\alpha }^{\dagger }+\text{H.c.}\right)
\notag \\
&&-\lambda \sum_{\mathbf{r}}\left( t_{\mathbf{r}\alpha }^{\dagger }
t_{\mathbf{r}\beta }+\text{H.c.}\right) ,  \label{ht}
\end{eqnarray}
where $\tilde{I}_{T}=I_{T}+I_{ST}$, and
$\tilde{I}_{T}^{p}=I_{T}^{p}-I_{ST}^{p}$.
To solve this problem analytically, we first consider the case $\lambda =0$.
After a Fourier transformation, the Hamiltonian maps onto
$H_{b}$
of Eq. (\ref{hb}) with $C=-3N\tilde{I}_{T}/4$, $\omega _{0k}=\tilde{I}_{T}$,
and $\omega _{1k}=-2\tilde{I}_{T}^{p}\gamma _{k}$. We then incorporate the last term using second-order perturbation theory.
From the results of the appendix
we obtain for the values of Table \ref{tab1} the ground state energy $\Delta
_{T}/N=-0.255$ meV for $\lambda =0$. The term $t_{\mathbf{r}\beta }^{\dagger
}t_{\mathbf{r}\alpha }$ applied to the ground state $|g\rangle $ (the vacuum
of Bogoliubov excitations $\alpha _{k}$ in the appendix) generates excited
states $|e_{k}\rangle =\alpha _{k}^{\dagger }t_{-k\beta }^{\dagger
}|g\rangle $ with an energy cost $\epsilon _{k}+\tilde{I}_{T}$. Therefore,
the usual correction to the energy in second-order
perturbation theory, using the results of the
appendix and calling $H_{\lambda }$ the last term of Eq. (\ref{ht}) takes
the form

\begin{equation}
\Delta _{T}^{c}=-\sum\limits_{k}\frac{|\langle e_{k}|H_{\lambda }|g\rangle
|^{2}}{\epsilon _{k}+\tilde{I}_{T}}
=-\sum_{k}\frac{\lambda ^{2}|v_{k}|^{2}}{\epsilon _{k}+\tilde{I}_{T}}.  \label{Deltac}
\end{equation}
The resulting value for this correction for the parameters of the model
becomes $\Delta _{T}^{c}/N=-0.011$ meV.

The small magnitude of this result may appear surprising, given that the maximum strength of the SOC is $\lambda/2=20$ meV and the gap is
$\tilde{I}_{T}=110.5$ eV. A naive estimate would suggest a first-order correction smaller but of the order of 20 meV. However, the first-order contribution vanishes for a pseudospin-singlet ground state, and the matrix elements connecting the ground state to the excited states are found to be very small. As a consequence, the second-order corrections are also small. Third-order contributions are expected to be further reduced by at least a factor
$\lambda/(2 \tilde{I}_{T})$ relative to the second-order ones, and therefore do not affect our conclusions.

For the Hamiltonian in the spin sector, using Eqs. (\ref{ham}) to
(\ref{dec}) and the fact that for the phase with orbital singlet dimers
$\left\langle
\mathbf{T}_{1\mathbf{r}}\cdot \mathbf{T}_{2\mathbf{r}}\right\rangle _{c}=-3/4
$, $\left\langle \mathbf{T}_{i\mathbf{r}}\right\rangle _{c}=\left\langle
T_{i\mathbf{r}}^{z}T_{i\mathbf{r}+\mathbf{a}}^{z}\right\rangle =0$, the
effective interplain interaction becomes $\tilde{I}_{S}/4=(I_{S}-3I_{ST})/4<0
$ and the intraplane one remains $I_{S}^{p}/4>0$. Using the symmetry
operation $R$ described above to rotate half of the spins, and introducing
the bosonic operators defined by Eq. (\ref{hol}) for both planes, the
Hamiltonian takes the form

\begin{eqnarray}
H_{S}^{\text{I}} &=&N\left( \frac{\tilde{I}_{S}-2I_{S}^{p}}{4}\right)
+\sum_{i,\mathbf{r}}\left( I_{S}^{p}-\frac{\tilde{I}_{S}}{4}\right)
b_{i\mathbf{r}}^{\dagger }b_{i\mathbf{r}}  \notag \\
&&+\sum_{\mathbf{r}}\frac{\tilde{I}_{S}}{4}\left( b_{1\mathbf{r}}^{\dagger
}b_{2\mathbf{r}}+\text{H.c.}\right)  \notag \\
&&+\sum_{i,\mathbf{r,a}}\frac{I_{S}^{p}}{4}\left( b_{i\mathbf{r}}^{\dagger
}b_{i\mathbf{r}+\mathbf{a}}^{\dagger }+\text{H.c.}\right) .  \label{hs}
\end{eqnarray}
Introducing the sum and the difference of the bosons for each vertical dimer
[note that the $s_{\mathbf{r}}$ below are different from the pseudospin
singlets of Eq. (\ref{cond})]

\begin{equation}
s_{\mathbf{r}}=\frac{b_{1\mathbf{r}}+b_{2\mathbf{r}}}{\sqrt{2}},
\text{ }d_{\mathbf{r}}=\frac{b_{1\mathbf{r}}-b_{2\mathbf{r}}}{\sqrt{2}},  \label{sd}
\end{equation}
and using the Fourier transform of the operators, $H_{S}^{\text{I}}$ is
reduced to the sum of two terms of the form of Eq. (\ref{hb}) solved in the
appendix

\begin{eqnarray}
H_{S}^{\text{I}} &=&N\left( \frac{\tilde{I}_{S}-2I_{S}^{p}}{4}\right)
+H_{s}+H_{d},  \notag \\
H_{s} &=&I_{S}^{p}\sum\limits_{k}s_{k}^{\dagger }s_{k}
+\frac{I_{S}^{p}}{2}\sum\limits_{k}\gamma _{k}\left( s_{k}^{\dagger }s_{-k}^{\dagger }
+\text{H.c.}\right) ,  \notag \\
H_{d} &=&\left( I_{S}^{p}-\frac{\tilde{I}_{S}}{2}\right)
\sum\limits_{k}d_{k}^{\dagger }d_{k}  \notag \\
&&+\frac{I_{S}^{p}}{2}\sum\limits_{k}\gamma _{k}\left( d_{k}^{\dagger
}d_{-k}^{\dagger }+\text{H.c.}\right) ,  \label{hsb}
\end{eqnarray}%
with $\omega _{0k}=I_{S}^{p}$ for $H_{s}$,
$\omega _{0k}=I_{S}^{p}-\tilde{I}_{S}/2$ for $H_{d}$,
and $\omega _{1k}=\tilde{I}_{S}\gamma _{k}$ in both cases.
Using the
results of the appendix and the parameters of Table \ref{tab1}, we obtain
for the respective corrections to the ground-sate energy $\Delta _{s}/N=-4.32
$ meV and $\Delta _{d}/N=-1.55$ meV. Adding the pseudospin contributions
$\Delta _{T}/N=-0.255$ meV and $\Delta _{T}^{c}/N=-0.011$ meV, we have a total contribution of -6.136 meV from spin fluctuations and triplets.
The remaining terms are the contribution of the pseudospin singlets $-3I_T/4=-51.075$ meV per site of the plane, the classical spin contribution
$I_S/4-I^p_S$
-53.65 meV, and the mixed spin-pseudospin term $-3I_{ST}/4=-31.8$ meV,
adding to -136.525 meV.
Thus, the total energy per atom of the plane of phase I
for the parameters of Table \ref{tab1} is

\begin{equation}
E_{g}(\text{I})=-142.661 \text{ meV.}  \label{ei}
\end{equation}

\subsubsection{Phase II}

\begin{figure}[ht]
\centering
\includegraphics[width=7.5cm]{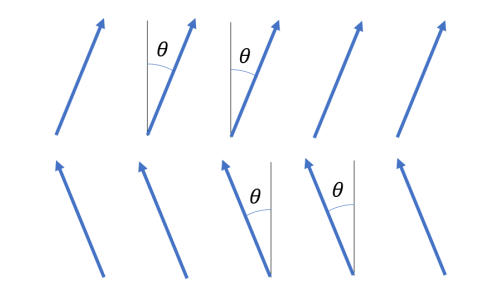}
\caption{(Color online) Classical order of the pseudospins in phase II
after the symmetry operation $R$ described in the text.}
\label{order}
\end{figure}

In this phase, one assumes $\left\langle
\mathbf{T}_{i\mathbf{r}}\right\rangle _{c}\neq 0$,
with long-range order of spins and pseudospins.
However, even at the classical level, the orientation of the pseudospin is
not trivial due to a competition between exchange and SOC terms. To
calculate the energy of this phase, we adopt a procedure similar to that of
Ref. \cite{Henry12}, obtaining first the classical order and then adding
quantum fluctuations on it. To have translational symmetry in the plane, we
apply again the symmetry operation $R$ described after Eq. (\ref{tbeta}),
which rotates spins and pseudospins in half of the sites an angle $\pi $
around the $x$ axis. Using as in phase I
$\left\langle \mathbf{\tilde{S}}_{1\mathbf{r}}\cdot \mathbf{\tilde{S}}_{2\mathbf{r}}\right\rangle
_{c}=\left\langle \mathbf{\tilde{S}}_{i\mathbf{r}}
\cdot \mathbf{\tilde{S}}_{i\mathbf{r}+\mathbf{a}}\right\rangle _{c}=\left\langle \tilde{S}_{i\mathbf{r}}^{z}\right\rangle _{c}=1$,
the Hamiltonian in the pseudospin sector takes
now the form

\begin{eqnarray}
H_{T}^{\text{II}} &=&\tilde{I}_{T}\sum_{\mathbf{r}}\mathbf{\tilde{T}}_{1\mathbf{r}}\cdot \mathbf{\tilde{T}}_{2\mathbf{r}}-\tilde{I}_{T}^{p}
\sum_{\mathbf{ra}}\tilde{T}_{i\mathbf{r}}^{z}\tilde{T}_{i\mathbf{r}+\mathbf{a}}^{z}
\notag \\
&&-\lambda \sum_{\mathbf{ra}}\tilde{T}_{i\mathbf{r}}^{y},  \notag \\
\tilde{I}_{T} &=&I_{T}+I_{ST},\text{ }\tilde{I}_{T}^{p}=I_{T}^{p}-I_{ST}^{p}.
\label{ht2}
\end{eqnarray}
Clearly, the last term prefers to have all pseudospins in the $y$ direction,
but to gain energy from the second term it is convenient to tilt the
pseudospins an angle $\theta $ in the $z$ direction (for positive
$\tilde{I}_{T}^{p}$ as in our case). Finally the first term favor opposite tilts in
both planes. Therefore the classical orientations of the spins are given by
(see Fig. \ref{order})

\begin{eqnarray}
\left\langle \tilde{T}_{i\mathbf{r}}^{x}\right\rangle _{c}
&=&0\text{, }\left\langle \tilde{T}_{i\mathbf{r}}^{y}\right\rangle _{c}=\frac{1}{2}\cos
\theta ,  \notag \\
\left\langle \tilde{T}_{1\mathbf{r}}^{z}\right\rangle _{c}
&=&\frac{1}{2}\sin \theta ,\text{ }\left\langle \tilde{T}_{2\mathbf{r}}^{z}\right\rangle
_{c}=-\frac{1}{2}\sin \theta ,  \label{clas}
\end{eqnarray}
leading to a classical energy per site of the plane

\begin{eqnarray}
E_{T}^{\text{II}} &=&\frac{\tilde{I}_{T}}{4}\cos (2\theta )
-\tilde{I}_{T}^{p}\sin ^{2}\theta   \notag \\
&&-\lambda \cos \theta .  \label{eclas}
\end{eqnarray}
Minimizing this energy one obtains $\theta =0$ if $\lambda \geq $
$\tilde{I}_{T}+2\tilde{I}_{T}^{p}$. However, in our case
$\lambda < \tilde{I}_{T}+2\tilde{I}_{T}^{p}$ and

\begin{equation}
\cos \theta =\frac{\lambda }{\tilde{I}_{T}+2\tilde{I}_{T}^{p}}.
\label{cosmin}
\end{equation}
For the parameters of Table \ref{tab1},
$\theta= 72.32 ^\circ$.

To introduce fluctuations, it is convenient to rotate the operators at each
site in such a way that the new local $z$ axis correspond to the direction
of the classical pseudospin. One way of doing this is to perform first a
$C_{3}$ rotation $y\rightarrow z\rightarrow x\rightarrow y$ and then a
rotation around the $y$ axis. After both operations, the transformed
operators become

\begin{eqnarray}
\hat{T}_{i\mathbf{r}}^{z} &=&\cos \theta
\tilde{T}_{i\mathbf{r}}^{y}-(-1)^{i}\sin \theta \tilde{T}_{i\mathbf{r}}^{z},  \notag \\
\hat{T}_{i\mathbf{r}}^{x} &=&(-1)^{i}\sin \theta
\tilde{T}_{i\mathbf{r}}^{y}+\cos \theta \tilde{T}_{i\mathbf{r}}^{z}.  \notag \\
\hat{T}_{i\mathbf{r}}^{y} &=&\tilde{T}_{i\mathbf{r}}^{x}.  \label{trans}
\end{eqnarray}
After a Holstein-Primakoff transformation of the new operators:
\begin{eqnarray}
\hat{T}_{i\mathbf{r}}^{z} &=&\frac{1}{2}
-a_{i\mathbf{r}}^{\dagger }a_{i\mathbf{r}},\text{ }\hat{T}_{i\mathbf{r}}^{x}=\frac{a_{i\mathbf{r}}^{\dagger
}+a_{i\mathbf{r}}}{2},  \notag \\
\hat{T}_{i\mathbf{r}}^{x} &=&i\frac{a_{i\mathbf{r}}-a_{i\mathbf{r}}^{\dagger
}}{2},  \label{hpl}
\end{eqnarray}
the linear terms in the bosonic operators vanish for $\theta $ satisfying
Eq. (\ref{cosmin}). Neglecting terms of higher order than two, defining
$s_{\mathbf{r}}=(a_{1\mathbf{r}}+a_{2\mathbf{r}})/\sqrt{2}$,
$d_{\mathbf{r}}=(a_{1\mathbf{r}}-b_{2\mathbf{r}})/\sqrt{2}$ [different from Eq. (\ref{sd})] and their Fourier transforms, the Hamiltonian takes the form

\begin{eqnarray}
H_{T}^{\text{II}} &=&NE_{T}^{\text{II}}+H_{T}^{s}+H_{T}^{d},  \notag \\
H_{T}^{s} &=&\sum_{k}[\omega _{0s}(k)s_{k}^{\dagger }s_{k}+\frac{\omega
_{1s}(k)}{2}(s_{k}^{\dagger }s_{-k}^{\dagger }+s_{k}s_{-k})],  \notag \\
H_{T}^{d} &=&\sum_{k}[\omega _{0d}(k)d_{k}^{\dagger }d_{k}+\frac{\omega
_{1d}(k)}{2}(d_{k}^{\dagger }d_{-k}^{\dagger }+d_{k}d_{-k})],  \label{htl}
\end{eqnarray}
with

\begin{eqnarray}
\omega _{0s}(k) &=&\tilde{I}_{T}\frac{\sin ^{2}\theta }{2}
+2\tilde{I}_{T}^{p}\sin ^{2}\theta +\lambda \cos \theta
-\frac{\tilde{I}_{T}^{p}}{4}\cos ^{2}\theta \,z\gamma _{k},  \notag \\
\omega _{1s}(k) &=&-\frac{\tilde{I}_{T}\sin ^{2}\theta }{2}
-\frac{\tilde{I}_{T}^{p}}{4}\cos ^{2}\theta \,z\gamma _{k},  \notag \\
\omega _{0d}(k) &=&\omega _{0,s}(k)-\tilde{I}_{T}\cos ^{2}\theta ,  \notag \\
\omega _{1d}(k) &=&\frac{\tilde{I}_{T}\sin ^{2}\theta }{2}
-\frac{\tilde{I}_{T}^{p}}{4}\cos ^{2}\theta \,z\gamma _{k}.  \label{ome}
\end{eqnarray}
$H_{T}^{s}$ and $H_{T}^{d}$ have the form of the bosonic Hamiltonian solved
in the appendix. The resulting energy gain due to fluctuations are $\Delta
_{T}^{s}/N=-3.27$ meV, $\Delta _{T}^{d}/N=-6.084$ meV.

The Hamiltonian in the spin sector has a similar form as for phase I.
However, for phase II,  $\left\langle \mathbf{\tilde{T}}_{1\mathbf{r}}\cdot
\mathbf{\tilde{T}}_{2\mathbf{r}}\right\rangle _{c}=\cos (2\theta )/4$,
$\left\langle T_{i\mathbf{r}}^{z}T_{i\mathbf{r}+\mathbf{a}}^{z}\right\rangle
_{c}=\sin^2\theta/4$, $\left\langle \tilde{T}_{i\mathbf{r}}^{y}\right\rangle _{c}=\cos
\theta /2$.
Therefore, from Eqs. (\ref{clas}) using Eqs. (\ref{ham}) to (\ref{dec}),  the effective interplain interaction becomes
$\tilde{I}_{S}^{\text{II}}/4=(I_{S}+\cos (2\theta )I_{ST})/4$ and the intraplane one
$\tilde{I}_{S}^{p}/4=(I_{S}^{p}+\sin ^{2}\theta I_{ST}^{p})/4$.  In
addition, there is a "magnetic field" $B=\cos (\theta ) \lambda /2$. Introducing the bosonic
operators defined by Eq. (\ref{hol}) for both planes, the Hamiltonian takes
the form

\begin{eqnarray}
H_{S}^{\text{II}} &=&NE_{S}^{\text{II}}
+\sum_{i,\mathbf{r}}\left( \tilde{I}_{S}^{p}-\frac{\tilde{I}_{S}^{\text{II}}}{4}
+ B \right) b_{i\mathbf{r}}^{\dagger }b_{i\mathbf{r}}  \notag \\
&&+\sum_{\mathbf{r}}\frac{\tilde{I}_{S}^{\text{II}}}{4}
\left( b_{1\mathbf{r}}^{\dagger }b_{2\mathbf{r}}+\text{H.c.}\right)   \notag \\
&&+\sum_{i,\mathbf{r,a}}\frac{\tilde{I}_{S}^{p}}{4}
\left( b_{i\mathbf{r}}^{\dagger }b_{i\mathbf{r}+\mathbf{a}}^{\dagger }+\text{H.c.}\right) ,
\label{hsl}
\end{eqnarray}
where

\begin{equation}
E_{S}^{\text{II}}=\frac{\tilde{I}_{S}^{\text{II}}
}{4}-\tilde{I}_{S}^{p} -2B ,
\label{ecls}
\end{equation}
is the classical energy per site of the plane.

Following the same procedure as in phase I, whereby symmetric and antisymmetric bosonic combinations are defined for each vertical dimer, $H_{S}^{\text{II}}$ takes the form

\begin{eqnarray}
H_{S}^{\text{II}} &=&NE_{S}^{\text{II}}+H_{s}^{\text{II}}+H_{d}^{\text{II}},
\notag \\
H_{s}^{\text{II}} &=&\left( \tilde{I}_{S}^{p}+\frac{\lambda }{2}\right)
\sum\limits_{k}s_{k}^{\dagger }s_{k}
+\frac{\tilde{I}_{S}^{p}}{2}\sum\limits_{k}\gamma _{k}\left( s_{k}^{\dagger }s_{-k}^{\dagger }
+\text{H.c.}\right) ,  \notag \\
H_{d}^{\text{II}} &=&\left( \tilde{I}_{S}^{p}+\frac{\lambda }{2}
-\frac{\tilde{I}_{S}^{\text{II}}}{2}\right) \sum\limits_{k}d_{k}^{\dagger }d_{k}
\notag \\
&&+\frac{\tilde{I}_{S}^{p}}{2}\sum\limits_{k}\gamma _{k}\left(
d_{k}^{\dagger }d_{-k}^{\dagger }+\text{H.c.}\right) ,  \label{hslb}
\end{eqnarray}
which has the form of the Hamiltonian of the appendix with
$\omega _{1k}=\tilde{I}_{S}^{p}\gamma _{k}$ in both cases and
$\omega _{0k}= \tilde{I}_{S}^{p}+\lambda /2$ for $H_{s}^{\text{II}}$ and
$\omega _{0k}=\tilde{I}_{S}^{p}+\lambda /2-\tilde{I}_{S}^{\text{II}}/2$ for $H_{d}^{\text{II}}$.
Using the results of the appendix and the parameters of
Table \ref{tab1}, we obtain for the respective corrections to
the ground-sate energy $\Delta_{S}^{s}/N=-3.012$  meV and
$\Delta _{S}^{d}/N=-1.8$ meV.

From Eq. (\ref{dec}), the total classical energy per site of the plane can be
written as

\begin{equation}
E_{c}^{\text{II}}=E_{T}^{\text{II}}+E_{S}^{\text{II}}-\frac{I_{ST}}{4}\cos (2\theta
)-I_{ST}^{p}\sin ^{2}\theta + \frac{\lambda}{2}\cos\theta.  \label{ec}
\end{equation}

For the values of Table \ref{tab1} we obtain that the sum of the last three terms
(those which couple spin and pseudospin) is 34.578 meV,
slightly larger than the mixed terms of phase I (31.8 meV)
Subtracting this
from $E_{T}^{\text{II}}$, one obtains that the pure pseudospin contribution
is -13.550 meV, smaller in magnitude than the gain of the pseudospin singlets in phase I ($-3I_T/4 \sim 51$ meV). Similarly the pure classical spin contribution
becomes -59.670 meV, slightly larger in magnitude that that
for phase I (-53.650 meV).
The total classical energy
becomes $E_{c}^{\text{II}}=$-103.969 meV. The total contribution of the quantum corrections
add to -14.166 meV (more than two times the contribution of the fluctuations
to phase I). This leads to the total energy of phase II

\begin{equation}
E_{g}(\text{II})=-118.135\text{ meV,}  \label{eii}
\end{equation}

This energy is larger than that of phase I by approximately 24 meV. The difference arises mainly from vertical interplane pseudospin fluctuations. More specifically, the energy difference between a singlet dimer and an antiferromagnetically ordered vertical dimer is $(I_T+I_{ST})/2 \sim 55$ meV. This contribution is partially compensated by the enhanced gain from intraplane pseudospin interactions (terms proportional to $I_{T}^{p}$), by an energy gain of nearly 9 meV due to the tilting of the pseudospins along the direction favoured by the SOC, and a larger contribution of spin and pseudospin fluctuations.
Nevertheless, phase I remains energetically favored.

This result is consistent with structural measurements, which detect neither lattice distortions nor displacements of the O atoms that would be expected if the pseudospin expectation value
$\left\langle T_{i\mathbf{r}}^{z}\right\rangle \neq 0$.

\section{Summary and discussion}
\label{sum}

The Ruddlesden-Popper series Sr$_{n+1}$Cr$_n$O$_{3n+1}$ consists of layers of $n$
CrO$_2$ planes with very weak interlayer connections. We have investigated
the effect of SOC on the layers with $n=1$ and $n=2$ starting from KKHs
without SOC derived earlier. The orbital degrees
of freedom are represented by pseudospin variables. The spins are treated by
spin variables and the pseudospin either by a similar approach or the BOF.

For the $n=1$ case, Sr$_2$CrO$_4$, we find that a realistic SOC
$\lambda=40$ meV dominates the physics, and the orbital degrees of freedom,
correspond to an entangled orbital phase for each
Cr site corresponding mainly to
either
$d_{xy\uparrow }^{\dagger }(d_{xz\uparrow }^{\dagger}
-id_{yz\uparrow }^{\dagger }) |0 \rangle$ or
$d_{xy\downarrow }^{\dagger}(d_{xz\downarrow }^{\dagger}
+id_{yz\downarrow }^{\dagger })|0 \rangle$
depending on the spin projection
$\langle S_{\mathbf{r}}^{z}\rangle$ at the corresponding site.
This result differs from the AF orbital order obtained in the absence of SOC \cite{Pandey21,Moha23} and is consistent with the recent findings of Mohapatra
{\it et al.}
\cite{Moha23}, who using a different approach and parameters,
find a transition to the SOC dominated phase at $\lambda =30$ meV.
The spin ordering is always the standard AF one.

For the $n=2$ compound, Sr$_3$Cr$_2$O$_7$, there is a competition between two phases in the KKH: I dominated by interplane orbital fluctuations, which favors pseudospin singlets of the form
$d_{1 xy \uparrow }^{\dagger} d_{2 xy \uparrow }^{\dagger }
(d_{1 xz \uparrow }^{\dagger} d_{2 yz \uparrow }^{\dagger }
-d_{1 yz \uparrow }^{\dagger} d_{2 yz \uparrow }^{\dagger })$ (or opposite spin)
for each two-dimensional position within the planes,
where 1 and 2 label the two planes.
Phase II corresponds to a state in which the pseudospin expectation value is nonzero at each site. This strongly suppresses vertical pseudospin fluctuations while allowing for a larger gain from antiferromagnetic intraplane
pseudospin interactions. As found earlier \cite{Ali19},in absence of spin–orbit coupling (SOC), vertical orbital fluctuations dominate the physics, and phase I has the lower energy. We find that SOC favors phase II. However, its effect is insufficient to make phase II the ground state.

One may wonder how sensitive these results are to the choice of parameters (see Table \ref{tab1}).
The value
$\lambda=40$ meV extracted from the atomic energy levels of Cr$^{4+}$ \cite{Moore}
is highly reliable and is not expected to vary significantly among different materials.
The rest of the parameters taken from Ref. \cite{Ali19}, depend on ratios of the form $t^2/\tilde{U}$, where $t$ is one of the effective Cr-Cr hoppings ($t_{z}=0.235$ eV for $xz,yz$ orbitals in the $z$ direction, $t_{p}=0.214$ eV for the $\alpha z$ orbitals in the $\alpha$ direction and $t_{xy}=0.248$ eV for the $xy$ orbitals in the plane)
and $\tilde{U}$ contains the on-site Coulomb repulsion $U \sim 4$ eV and other
intrasite Cr
interactions proportional to the exchange Hund coupling $J$.
We believe that $t$ obtained by maximally localized-Wannier functions
calculated by DFT are also robust. In contrast, $U$ depends on the particular
screening and can change among materials, whereas $J \sim 0.7$ eV determined
from atomic values is not expected to change considerably \cite{Ali13}.
Increasing (decreasing) the value of $U$ leads to a decrease (enhancement) in the interaction parameters of the KKH, thereby making the effects of SOC more (less) important.

Therefore, for Sr$_2$CrO$_4$, an increase in $U$ does not modify our conclusions.
For the parameters of Table \ref{tab1}, the critical value of $\lambda$
to reach the SOC dominated phase is $\sim 16$ meV. If all intrasite
Cr interactions
were decreased by a factor two, this critical value would double, yet it would still remain well below the actual value of $\lambda$.

A previous Hartree-Fock study of the three-band Hubbard model without
SOC suggests that, for our ratio $J/U=0.175$, decreasing $U$ drives a transition from the AF orbital phase to a staggered-orbital-stripe phase (SOS) for $U \sim$ 1.8 eV \cite{Pandey21}.
We have not explicitly investigated the SOS phase. However, the inclusion of SOC destabilizes not only the AF orbital phase but also the SOS phase. Therefore, our result for an entangled orbital phase remains robust even
for such small value of $U$.

In contrast, for Sr$_3$Cr$_2$O$_7$ a reduction of the on-site Cr interactions does not affect our conclusions. Moreover, we find that increasing $U$ to 6 eV likewise does not modify our results.
As for the single-layer compound, we have not explored more complex phases, such as those shown in Fig. 6 of the Supplementary Material of Ref. \cite{Jean17}, which were obtained from exact diagonalization of small clusters. However, these calculations were performed for interaction strengths much smaller than realistic values. Furthermore, only the phases we consider are consistent with the experimentally observed magnetic structure \cite{Jean17}.

Our calculations were performed at zero temperature. Based on the finite-temperature phase diagram of the transverse Ising model \cite{Friedman78}, we expect that, for the single-layer compound, the entangled orbital phase is further stabilized as temperature increases. For the bilayer compound, phase II is likewise expected to be favored at higher temperatures due to the larger number of low-energy excitations. Nevertheless, since the energy difference with phase I (characterised by orbital singlets) is on the order of 24 meV, moderate temperatures cannot affect our conclusions.

Our results are consistent with the observed AF spin ordering in the single-layer compound \cite{Sugi14,Jean19}, as well as with AF spin order within the planes and ferromagnetic between planes in the bilayer compound \cite{Jean17}. In addition, the presence of long-range orbital ordering would be expected to induce structural distortions that reflect the resulting anisotropy of the electronic structure (see the end of Section \ref{one}). However, no such distortions have been observed. Our results for both compounds namely, an entangled orbital state in the single-layer system and out-of-plane singlet dimers in the bilayer) are consistent with this absence of distortions. We therefore conclude that our findings are in good agreement with experiment for both compounds.

\section*{Acknowledgments}

C. F. has a scholarship of Instituto Balseiro.

\appendix

\section{Diagonalization of the bosonic Hamiltonian}

\label{app}

The Hamiltonian $H_{b}$ given by Eq. (\ref{hb}) and reproduced below for
completeness

\begin{equation}
H_{b}=C+\sum_{k}\left[ \omega _{0k}b_{k}^{\dagger }b_{k}+\left( \frac{\omega
_{1k}}{2}b_{k}^{\dagger }b_{-k}^{\dagger }+\text{H.c.}\right) \right] ,
\label{hb2}
\end{equation}
can be transformed into the diagonal form
\begin{equation}
H_{B}=C+\Delta +\sum_{k}\epsilon _{k}\alpha _{k}^{\dagger }\alpha _{k}
\label{diag}
\end{equation}
by means of a Bogoliubov transformation

\begin{equation}
\alpha _{k}=u_{k}b_{k}-v_{k}b_{-k}^{\dagger },  \label{alpha}
\end{equation}
where $u_{k}$ can be chosen real and positive, and both coefficients should
satisfy the relation

\begin{equation}
\left[ \alpha _{k},\alpha _{k}^{\dagger }\right] =u_{k}^{2}-|v_{k}|^{2}=1.
\label{conm}
\end{equation}
Using this equation one can invert Eq. (\ref{alpha})

\begin{equation}
b_{k}=u_{k}\alpha _{k}+v_{-k}\alpha _{-k}^{\dagger }.  \label{inver}
\end{equation}
From the equation $\left[ \alpha _{k},H_{B}\right]
=\epsilon _{k}\alpha _{k}$, one obtains the energies of the quasiparticles and the coefficients:

\begin{eqnarray}
\epsilon _{k} &=&\sqrt{\omega _{0k}^{2}-|\omega _{1k}|^{2}},
\text{ }u_{k}=\sqrt{\frac{1}{2}+\frac{\omega _{0k}}{2\epsilon _{k}}},  \notag \\
v_{k} &=&\frac{-u_{k}\omega _{1k}}{\epsilon _{k}+\omega _{0k}}.  \label{ek}
\end{eqnarray}
Using Eq. (\ref{inver}) one obtains for the state without quasiparticles

\begin{equation}
\left\langle b_{k}^{\dagger }b_{k}\right\rangle =|v_{-k}|^{2}=\frac{\omega _{0-k}}{2\epsilon _{-k}}-\frac{1}{2}
\text{, }\left\langle b_{k}^{\dagger }b_{-k}^{\dagger }\right\rangle
=\bar{v}_{-k}u_{-k}.  \label{expe}
\end{equation}
Replacing this expression in the expectation value of $H_{B}$, one obtains
for the correction to the ground state energy in the usual case in which
$\omega _{0-k}=\omega _{0k}$

\begin{equation}
\Delta =-\sum_{k}\epsilon _{k}|v_{k}|^{2}=\sum_{k}\frac{\epsilon _{k}-\omega
_{0k}}{2}.  \label{ene}
\end{equation}
The total ground-state energy is $C+\Delta $.

\bibliography{ref}

@PREAMBLE{
 "\providecommand{\noopsort}[1]{}" 
 # "\providecommand{\singleletter}[1]{#1}%" 
}

@article{Kugel82,
doi = {10.1070/PU1982v025n04ABEH004537},
url = {https://doi.org/10.1070/PU1982v025n04ABEH004537},
year = {1982},
month = {apr},
publisher = {},
volume = {25},
number = {4},
pages = {231},
author = {Kliment I Kugel' and D I Khomskiĭ},
title = {{The Jahn-Teller effect and magnetism: transition metal
compounds}},
journal = {Soviet Physics Uspekhi}
}

@article{Liech95,
  title = {Density-functional theory and strong interactions: Orbital ordering in Mott-Hubbard insulators},
  author = {Liechtenstein, A. I. and Anisimov, V. I. and Zaanen, J.},
  journal = {Phys. Rev. B},
  volume = {52},
  issue = {8},
  pages = {R5467--R5470},
  numpages = {0},
  year = {1995},
  month = {Aug},
  publisher = {American Physical Society},
  doi = {10.1103/PhysRevB.52.R5467},
  url = {https://link.aps.org/doi/10.1103/PhysRevB.52.R5467}
}

@article{Mizo95,
  title = {Unrestricted Hartree-Fock study of transition-metal oxides: Spin and orbital ordering in perovskite-type lattice},
  author = {Mizokawa, T. and Fujimori, A.},
  journal = {Phys. Rev. B},
  volume = {51},
  issue = {18},
  pages = {12880--12883},
  numpages = {0},
  year = {1995},
  month = {May},
  publisher = {American Physical Society},
  doi = {10.1103/PhysRevB.51.12880},
  url = {https://link.aps.org/doi/10.1103/PhysRevB.51.12880}
}

@article{Feiner97,
  title = {Quantum Melting of Magnetic Order due to Orbital Fluctuations},
  author = {Feiner, Louis Felix and Ole\ifmmode \acute{s}\else \'{s}\fi{}, Andrzej M. and Zaanen, Jan},
  journal = {Phys. Rev. Lett.},
  volume = {78},
  issue = {14},
  pages = {2799--2802},
  numpages = {0},
  year = {1997},
  month = {Apr},
  publisher = {American Physical Society},
  doi = {10.1103/PhysRevLett.78.2799},
  url = {https://link.aps.org/doi/10.1103/PhysRevLett.78.2799}
}

@article{Ul03,
  title = {{Magnetic Neutron Scattering Study of ${\mathrm{Y}\mathrm{V}\mathrm{O}}_{3}$: Evidence for an Orbital Peierls State}},
  author = {Ulrich, C. and Khaliullin, G. and Sirker, J. and Reehuis, M. and Ohl, M. and Miyasaka, S. and Tokura, Y. and Keimer, B.},
  journal = {Phys. Rev. Lett.},
  volume = {91},
  issue = {25},
  pages = {257202},
  numpages = {4},
  year = {2003},
  month = {Dec},
  publisher = {American Physical Society},
  doi = {10.1103/PhysRevLett.91.257202},
  url = {https://link.aps.org/doi/10.1103/PhysRevLett.91.257202}
}

@article{Ali04,
  title = {{Magnetic and orbital ordering of ${\mathrm{RuO}}_{2}$ planes in ${\mathrm{RuSr}}_{2}(\mathrm{Eu},\mathrm{Gd}){\mathrm{Cu}}_{2}{\mathrm{O}}_{8}$}},
  author = {Aligia, A. A. and Gusm\~ao, M. A.},
  journal = {Phys. Rev. B},
  volume = {70},
  issue = {5},
  pages = {054403},
  numpages = {7},
  year = {2004},
  month = {Aug},
  publisher = {American Physical Society},
  doi = {10.1103/PhysRevB.70.054403},
  url = {https://link.aps.org/doi/10.1103/PhysRevB.70.054403}
}

@article{Sugai06,
  title = {{Orbital waves in $\mathrm{Y}\mathrm{V}{\mathrm{O}}_{3}$ studied by Raman scattering}},
  author = {Sugai, S. and Hirota, K.},
  journal = {Phys. Rev. B},
  volume = {73},
  issue = {2},
  pages = {020409},
  numpages = {4},
  year = {2006},
  month = {Jan},
  publisher = {American Physical Society},
  doi = {10.1103/PhysRevB.73.020409},
  url = {https://link.aps.org/doi/10.1103/PhysRevB.73.020409}
}

@Article{Lee06,
author={Lee, Seongsu
and Park, J.-G.
and Adroja, D. T.
and Khomskii, D.
and Streltsov, S.
and McEwen, K. A.
and Sakai, H.
and Yoshimura, K.
and Anisimov, V. I.
and Mori, D.
and Kanno, R.
and Ibberson, R.},
title={{Spin gap in ${\mathrm{Tl}_{3}\mathrm{Ru}_{2}\mathrm{O}_{7}}$ and the possible formation of Haldane chains in three-dimensional crystals}},
journal={Nature Materials},
year={2006},
month={Jun},
day={01},
volume={5},
number={6},
pages={471-476},
issn={1476-4660},
doi={10.1038/nmat1605},
url={https://doi.org/10.1038/nmat1605}
}

@article{Ray07,
  title = {{Orbital Fluctuations in the Different Phases of ${\mathrm{LaVO}}_{3}$ and ${\mathrm{YVO}}_{3}$}},
  author = {De Raychaudhury, M. and Pavarini, E. and Andersen, O. K.},
  journal = {Phys. Rev. Lett.},
  volume = {99},
  issue = {12},
  pages = {126402},
  numpages = {4},
  year = {2007},
  month = {Sep},
  publisher = {American Physical Society},
  doi = {10.1103/PhysRevLett.99.126402},
  url = {https://link.aps.org/doi/10.1103/PhysRevLett.99.126402}
}

@article{Manaka07,
author = {Manaka ,Hirotaka and Miyashita ,Yusuke and Watanabe ,Yusuke and Masuda ,Takatsugu},
title = {{Synthesis of Double-Layer Perovskite Fluoride ${\mathrm{K}_{3}\mathrm{Cu}_{2}\mathrm{F}_{7}}$ with Spin Gap and Orbital Order}},
journal = {Journal of the Physical Society of Japan},
volume = {76},
number = {4},
pages = {044710},
year = {2007},
doi = {10.1143/JPSJ.76.044710},

URL = { 
    
        https://doi.org/10.1143/JPSJ.76.044710

},
eprint = { 
    
        https://doi.org/10.1143/JPSJ.76.044710
    
}
}

@article{Normand08,
  title = {Frustration and entanglement in the ${t}_{2g}$ spin-orbital model on a triangular lattice: Valence-bond and generalized liquid states},
  author = {Normand, Bruce and Ole\ifmmode \acute{s}\else \'{s}\fi{}, Andrzej M.},
  journal = {Phys. Rev. B},
  volume = {78},
  issue = {9},
  pages = {094427},
  numpages = {39},
  year = {2008},
  month = {Sep},
  publisher = {American Physical Society},
  doi = {10.1103/PhysRevB.78.094427},
  url = {https://link.aps.org/doi/10.1103/PhysRevB.78.094427}
}

@article{Kata09,
  title = {{Anomalous Metallic State in the Vicinity of Metal to Valence-Bond Solid Insulator Transition in ${\mathrm{LiVS}}_{2}$}},
  author = {Katayama, N. and Uchida, M. and Hashizume, D. and Niitaka, S. and Matsuno, J. and Matsumura, D. and Nishihata, Y. and Mizuki, J. and Takeshita, N. and Gauzzi, A. and Nohara, M. and Takagi, H.},
  journal = {Phys. Rev. Lett.},
  volume = {103},
  issue = {14},
  pages = {146405},
  numpages = {4},
  year = {2009},
  month = {Oct},
  publisher = {American Physical Society},
  doi = {10.1103/PhysRevLett.103.146405},
  url = {https://link.aps.org/doi/10.1103/PhysRevLett.103.146405}
}

@article{Stin11,
  title = {{Symmetry-Breaking Lattice Distortion in ${\mathrm{Sr}}_{3}{\mathrm{Ru}}_{2}{\mathrm{O}}_{7}$}},
  author = {Stingl, C. and Perry, R. S. and Maeno, Y. and Gegenwart, P.},
  journal = {Phys. Rev. Lett.},
  volume = {107},
  issue = {2},
  pages = {026404},
  numpages = {4},
  year = {2011},
  month = {Jul},
  publisher = {American Physical Society},
  doi = {10.1103/PhysRevLett.107.026404},
  url = {https://link.aps.org/doi/10.1103/PhysRevLett.107.026404}
}

@article{Nun13,
  title = {{Orbital Kondo effect in V-doped $1T$-CrSe${}_{2}$}},
  author = {N\'u\~nez, Mat\'{\i}as and Freitas, Daniele C. and Gay, Fr\'ed\'erique and Marcus, Jacques and Strobel, Pierre and Aligia, Armando A. and N\'u\~nez-Regueiro, Manuel},
  journal = {Phys. Rev. B},
  volume = {88},
  issue = {24},
  pages = {245129},
  numpages = {11},
  year = {2013},
  month = {Dec},
  publisher = {American Physical Society},
  doi = {10.1103/PhysRevB.88.245129},
  url = {https://link.aps.org/doi/10.1103/PhysRevB.88.245129}
}

@article{Lobos14,
doi = {10.1088/1742-6596/568/5/052002},
url = {https://doi.org/10.1088/1742-6596/568/5/052002},
year = {2014},
month = {dec},
publisher = {},
volume = {568},
number = {5},
pages = {052002},
author = {Lobos, A M and Aligia, A A},
title = {{Magnetic and orbital instabilities in a lattice of SU(4) organometallic Kondo complexes}},
journal = {Journal of Physics: Conference Series}
}

@article{Gio14,
  title = {{Cooperative effects of Jahn-Teller distortion, magnetism, and Hund's coupling in the insulating phase of ${\mathrm{BaCrO}}_{3}$}},
  author = {Giovannetti, Gianluca and Aichhorn, Markus and Capone, Massimo},
  journal = {Phys. Rev. B},
  volume = {90},
  issue = {24},
  pages = {245134},
  numpages = {6},
  year = {2014},
  month = {Dec},
  publisher = {American Physical Society},
  doi = {10.1103/PhysRevB.90.245134},
  url = {https://link.aps.org/doi/10.1103/PhysRevB.90.245134}
}

@article{Ishi17,
author = {Ishikawa ,Takashi and Toriyama ,Tatsuya and Konishi ,Takehisa and Sakurai ,Hiroya and Ohta ,Yukinori},
title = {{Reversed Crystal-Field Splitting and Spin–Orbital Ordering in ${\alpha-\mathrm{Sr}}_{2}{\mathrm{Cr}}{\mathrm{O}}_{4}$}},
journal = {Journal of the Physical Society of Japan},
volume = {86},
number = {3},
pages = {033701},
year = {2017},
doi = {10.7566/JPSJ.86.033701},
URL = {https://doi.org/10.7566/JPSJ.86.033701}
}

@article{Ortega07,
  title = {{Microstrain Sensitivity of Orbital and Electronic Phase Separation in ${\mathrm{SrCrO}}_{3}$}},
  author = {Ortega-San-Martin, Luis and Williams, Anthony J. and Rodgers, Jennifer and Attfield, J. Paul and Heymann, Gunter and Huppertz, Hubert},
  journal = {Phys. Rev. Lett.},
  volume = {99},
  issue = {25},
  pages = {255701},
  numpages = {4},
  year = {2007},
  month = {Dec},
  publisher = {American Physical Society},
  doi = {10.1103/PhysRevLett.99.255701},
  url = {https://link.aps.org/doi/10.1103/PhysRevLett.99.255701}
}

@article{Lee09,
  title = {{Orbital-ordering driven structural distortion in metallic ${\text{SrCrO}}_{3}$}},
  author = {Lee, K.-W. and Pickett, W. E.},
  journal = {Phys. Rev. B},
  volume = {80},
  issue = {12},
  pages = {125133},
  numpages = {6},
  year = {2009},
  month = {Sep},
  publisher = {American Physical Society},
  doi = {10.1103/PhysRevB.80.125133},
  url = {https://link.aps.org/doi/10.1103/PhysRevB.80.125133}
}

@article{Jean17,
  title = {{Singlet Orbital Ordering in Bilayer ${\mathrm{Sr}}_{3}{\mathrm{Cr}}_{2}{\mathrm{O}}_{7}$}},
  author = {Jeanneau, Justin and Toulemonde, Pierre and Remenyi, Gyorgy and Sulpice, Andr\'e and Colin, Claire and Nassif, Vivian and Suard, Emmanuelle and Salas Colera, Eduardo and Castro, Germ\'an R. and Gay, Frederic and Urdaniz, Corina and Weht, Ruben and Fevrier, Clement and Ralko, Arnaud and Lacroix, Claudine and Aligia, Armando A. and N\'u\~nez-Regueiro, Manuel},
  journal = {Phys. Rev. Lett.},
  volume = {118},
  issue = {20},
  pages = {207207},
  numpages = {5},
  year = {2017},
  month = {May},
  publisher = {American Physical Society},
  doi = {10.1103/PhysRevLett.118.207207},
  url = {https://link.aps.org/doi/10.1103/PhysRevLett.118.207207}
}

@article{Jean19,
doi = {10.1209/0295-5075/127/27002},
url = {https://doi.org/10.1209/0295-5075/127/27002},
year = {2019},
month = {sep},
publisher = {EDP Sciences, IOP Publishing and Società Italiana di Fisica},
volume = {127},
number = {2},
pages = {27002},
author = {Jeanneau, Justin and Toulemonde, Pierre and Remenyi, Gyorgy and Sulpice, André and Colin, Claire V. and Nassif, Vivian and Suard, Emmanuelle and Gay, Frederic and Weht, Ruben and Núñez-Regueiro, Manuel},
title = {{Magnetism and anomalous apparently inverse Jahn-Teller effect in ${\mathrm{Sr}}_{2}{\mathrm{Cr}}{\mathrm{O}}_{4}$}},
journal = {Europhysics Letters}
}

@article{Ali19,
  title = {{Spin and orbital ordering in bilayer ${\mathrm{Sr}}_{3}{\mathrm{Cr}}_{2}{\mathrm{O}}_{7}$}},
  author = {Aligia, Armando A. and Helman, Christian},
  journal = {Phys. Rev. B},
  volume = {99},
  issue = {19},
  pages = {195150},
  numpages = {8},
  year = {2019},
  month = {May},
  publisher = {American Physical Society},
  doi = {10.1103/PhysRevB.99.195150},
  url = {https://link.aps.org/doi/10.1103/PhysRevB.99.195150}
}

@article{Pandey21,
  title = {{Origin of the magnetic and orbital ordering in $\ensuremath{\alpha}\text{\ensuremath{-}}{\mathrm{Sr}}_{2}\mathrm{Cr}{\mathrm{O}}_{4}$}},
  author = {Pandey, Bradraj and Zhang, Yang and Kaushal, Nitin and Soni, Rahul and Lin, Ling-Fang and Hu, Wen-Jun and Alvarez, Gonzalo and Dagotto, Elbio},
  journal = {Phys. Rev. B},
  volume = {103},
  issue = {4},
  pages = {045115},
  numpages = {11},
  year = {2021},
  month = {Jan},
  publisher = {American Physical Society},
  doi = {10.1103/PhysRevB.103.045115},
  url = {https://link.aps.org/doi/10.1103/PhysRevB.103.045115}
}

@article{Carta22,
  title = {{Evidence for Jahn-Teller-driven metal-insulator transition in strained ${\mathrm{SrCrO}}_{3}$ from first-principles calculations}},
  author = {Carta, Alberto and Ederer, Claude},
  journal = {Phys. Rev. Mater.},
  volume = {6},
  issue = {7},
  pages = {075004},
  numpages = {8},
  year = {2022},
  month = {Jul},
  publisher = {American Physical Society},
  doi = {10.1103/PhysRevMaterials.6.075004},
  url = {https://link.aps.org/doi/10.1103/PhysRevMaterials.6.075004}
}

@article{Moha23,
doi = {10.1088/1361-648X/ace872},
url = {https://doi.org/10.1088/1361-648X/ace872},
year = {2023},
month = {jul},
publisher = {IOP Publishing},
volume = {35},
number = {43},
pages = {435601},
author = {Mohapatra, Shubhajyoti and Kumar Singh, Dheeraj and Singh, Avinash},
title = {{Spin–orbit coupling and magnetism in ${\mathrm{Sr}}_{2}{\mathrm{Cr}}{\mathrm{O}}_{4}$}},
journal = {Journal of Physics: Condensed Matter}
}

@article{Doyle24,
  title = {{Effects of dimensionality on the electronic structure of Ruddlesden-Popper chromates ${\mathrm{Sr}}_{n+1}{\mathrm{Cr}}_{n}{\mathrm{O}}_{3n+1}$}},
  author = {Doyle, Spencer and Takana, Lerato and Anderson, Margaret A. and Ferenc Segedin, Dan and El-Sherif, Hesham and Brooks, Charles M. and Wang, Xiaoping and Shafer, Padraic and N'Diaye, Alpha T. and Baggari, Ismail El and Ratcliff, William D. and Cano, Andr\'es and Meier, Quintin N. and Mundy, Julia A.},
  journal = {Phys. Rev. Mater.},
  volume = {8},
  issue = {7},
  pages = {L071602},
  numpages = {10},
  year = {2024},
  month = {Jul},
  publisher = {American Physical Society},
  doi = {10.1103/PhysRevMaterials.8.L071602},
  url = {https://link.aps.org/doi/10.1103/PhysRevMaterials.8.L071602}
}

@Article{Chiz25,
author={Chizhov, D. E.
and Igoshev, P. A.},
title={Study of the Ground State of the Three-Orbital Model of Layered Perovskite},
journal={Bulletin of the Russian Academy of Sciences: Physics},
year={2025},
month={Nov},
day={01},
volume={89},
number={11},
pages={2134-2139},
issn={1934-9432},
doi={10.1134/S1062873825713297},
url={https://doi.org/10.1134/S1062873825713297}
}

@article{Meier26,
  title = {{Net and Compensated Altermagnetism from Staggered Orbital Order: Layer-Dependent Spin Splitting in ${\mathrm{Sr}}_{n+1}{\mathrm{Cr}}_{n}{\mathrm{O}}_{3n+1}$}},
  author = {Meier, Quintin N. and Carta, Alberto and Ederer, Claude and Cano, Andr\'es},
  journal = {Phys. Rev. Lett.},
  volume = {136},
  issue = {11},
  pages = {116705},
  numpages = {8},
  year = {2026},
  month = {Mar},
  publisher = {American Physical Society},
  doi = {10.1103/mm8t-82q4},
  url = {https://link.aps.org/doi/10.1103/mm8t-82q4}
}

@article{Ali13,
  title = {{Effect of covalency and interactions on the trigonal splitting in Na${}_{x}$CoO${}_{2}$}},
  author = {Aligia, A. A.},
  journal = {Phys. Rev. B},
  volume = {88},
  issue = {7},
  pages = {075128},
  numpages = {5},
  year = {2013},
  month = {Aug},
  publisher = {American Physical Society},
  doi = {10.1103/PhysRevB.88.075128},
  url = {https://link.aps.org/doi/10.1103/PhysRevB.88.075128}
}

@article{Muniz14,
    author = {Muniz, Rodrigo A. and Kato, Yasuyuki and Batista, Cristian D.},
    title = {Generalized spin-wave theory: Application to the bilinear–biquadratic model},
    journal = {Progress of Theoretical and Experimental Physics},
    volume = {2014},
    number = {8},
    pages = {083I01},
    year = {2014},
    month = {08},
    issn = {2050-3911},
    doi = {10.1093/ptep/ptu109},
    url = {https://doi.org/10.1093/ptep/ptu109}
}

@article{Gopa94,
  title = {Spin ladders with spin gaps: A description of a class of cuprates},
  author = {Gopalan, Sudha and Rice, T. M. and Sigrist, M.},
  journal = {Phys. Rev. B},
  volume = {49},
  issue = {13},
  pages = {8901--8910},
  numpages = {0},
  year = {1994},
  month = {Apr},
  publisher = {American Physical Society},
  doi = {10.1103/PhysRevB.49.8901},
  url = {https://link.aps.org/doi/10.1103/PhysRevB.49.8901}
}

@article{Matsu04,
  title = {{Field- and pressure-induced magnetic quantum phase transitions in ${\mathrm{TlCuCl}}_{3}$}},
  author = {Matsumoto, Masashige and Normand, B. and Rice, T. M. and Sigrist, Manfred},
  journal = {Phys. Rev. B},
  volume = {69},
  issue = {5},
  pages = {054423},
  numpages = {20},
  year = {2004},
  month = {Feb},
  publisher = {American Physical Society},
  doi = {10.1103/PhysRevB.69.054423},
  url = {https://link.aps.org/doi/10.1103/PhysRevB.69.054423}
}

@article{Nor11,
  title = {Complete bond-operator theory of the two-chain spin ladder},
  author = {Normand, B. and R\"uegg, Ch.},
  journal = {Phys. Rev. B},
  volume = {83},
  issue = {5},
  pages = {054415},
  numpages = {12},
  year = {2011},
  month = {Feb},
  publisher = {American Physical Society},
  doi = {10.1103/PhysRevB.83.054415},
  url = {https://link.aps.org/doi/10.1103/PhysRevB.83.054415}
}

@article{Schmitt22,
  author    = {Markus Schmitt and Marek M. Rams and Jacek Dziarmaga and Markus Heyl and Wojciech H. Zurek},
  title     = {{Quantum phase transition dynamics in the two-dimensional transverse-field Ising model}},
  journal   = {Science Advances},
  volume    = {8},
  number    = {37},
  pages     = {eabl6850},
  doi       = {10.1126/sciadv.abl6850},
  year      = {2022},
}

@article{Friedman78,
  title={{Ising model with a transverse field in two dimensions: Phase diagram and critical properties from a real-space renormalization group}},
  author={Friedman, Zvi},
  journal={Physical Review B},
  volume={17},
  number={3},
  pages={1429},
  year={1978},
  publisher={APS}
}

@book{Moore,
  title={{Atomic Energy Levels. Vol. 2}},
  author={Moore, Charlotte E},
  year={1959},
  publisher={US Governm. Print. Office}
}

@article{Blote02,
  title = {{Cluster Monte Carlo simulation of the transverse Ising model}},
  author = {Bl\"ote, Henk W. J. and Deng, Youjin},
  journal = {Phys. Rev. E},
  volume = {66},
  issue = {6},
  pages = {066110},
  numpages = {8},
  year = {2002},
  month = {Dec},
  publisher = {American Physical Society},
  doi = {10.1103/PhysRevE.66.066110},
  url = {https://link.aps.org/doi/10.1103/PhysRevE.66.066110}
}

@article{Henry12,
  title={{Spin-wave analysis of the transverse-field Ising model on the checkerboard lattice}},
  author={Henry, Louis-Paul and Holdsworth, Peter CW and Mila, Fr{\'e}d{\'e}ric and Roscilde, Tommaso},
  journal={Physical Review B—Condensed Matter and Materials Physics},
  volume={85},
  number={13},
  pages={134427},
  year={2012},
  publisher={APS}
}

@article{Sugi14,
doi = {10.1088/1742-6596/551/1/012011},
url = {https://doi.org/10.1088/1742-6596/551/1/012011},
year = {2014},
month = {dec},
publisher = {},
volume = {551},
number = {1},
pages = {012011},
author = {Sugiyama, J and Nozaki, H and Umegaki, I and Higemoto, W and Ansaldo, E J and Brewer, J H and Sakurai, H and Kao, T-H and Yang, H-D and Månsson, M},
title = {{Microscopic magnetic nature of K$_2$NiF$_4$-type 3d transition metal oxides}},
journal = {Journal of Physics: Conference Series}
}

@article{Yama19,
  title = {{Contrasting Pressure-Induced Metallization Processes in Layered Perovskites, $\ensuremath{\alpha}$-${\mathrm{Sr}}_{2}M{\mathrm{O}}_{4}$ ($M=\mathrm{V}$, Cr)}},
  author = {Yamauchi, Touru and Shimazu, Taku and Nishio-Hamane, Daisuke and Sakurai, Hiroya},
  journal = {Phys. Rev. Lett.},
  volume = {123},
  issue = {15},
  pages = {156601},
  numpages = {5},
  year = {2019},
  month = {Oct},
  publisher = {American Physical Society},
  doi = {10.1103/PhysRevLett.123.156601},
  url = {https://link.aps.org/doi/10.1103/PhysRevLett.123.156601}
}

\end{document}